\documentclass[amssymb,prd,twocolumn,superscriptaddress,aps,nofootinbib,showpacs]{revtex4-1}
\usepackage{graphicx, epsfig, amssymb} 
\usepackage{amsmath, amsfonts}
\usepackage{bm} 
\usepackage[breaklinks]{hyperref}
\usepackage{color}
 \usepackage[usenames]{xcolor}

\newcommand{\be}{\begin{equation}}
\newcommand{\ee}{\end{equation}}                  
\newcommand{\bea}{\begin{eqnarray}}
\newcommand{\eea}{\end{eqnarray}}

\begin{document}
\title{I-Love irrotationally}
\author{T\'{e}rence Delsate}
\email{terence.delsate(at)umons.ac.be}
\affiliation{Theoretical and Mathematical Physics Dept.,\\
Universit\'{e} de Mons - UMONS, 20, Place du Parc, 7000 Mons, Belgium}

\date{\today}

\begin{abstract}
In this short note, we investigate the existence of universal relations between the gravimagnetic Love number of irrotational stars and the dimensionless moment of inertia. These Love numbers take into account the internal motion of the fluid, while the star is globally irrotational. The goal is to extend the so-called I-Love-Q relations - providing a strong correlation between the gravitoelectric Love number, the dimensionless moment of inertia and the dimensionless rotation-induced quadrupole - to the gravitomagnetic sector, where internal motion is taken into account. As a byproduct, we present for the first time this new gravitomagnetic Love number for realistic equations of state.
\end{abstract}

\maketitle

\section{Introduction}

Love numbers encode the deformation of self-gravitating fluids towards external tidal perturbations. While tidal interactions have been well understood for more than a century in the Newtonian context, \cite{Love:1911}, their relativistic counterpart has only started being understood recently. First studies investigated the adiabatic limit of static configurations \cite{Damour:Nagar:2009:4, Binnington:Poisson:2009}, i.e. under the assumption that the external tidal field varies slowly. Recently, the formalism has been extended to slowly rotating configurations \cite{Pani:2015hfa, Landry:2015zfa}, still in the adiabatic limit. A formalism to extend the tidal response beyond the adiabatic limit, relying on effective field theory has been introduced in the Newtonian limit in \cite{Chakrabarti:2013xza} and extended to the relativistic case in \cite{Chakrabarti:2013lua}.

In the relativistic case, static configurations are affected by two sector of gravitational perturbations: odd and even, corresponding to gravitomagnetic and gravitoelectric perturbations. The gravitomagnetic sector is particularly relevant to the case where rotation is present. This is a purely relativistic effect and doesn't have a Newtonian analogue: a central body surrounded by an orbiting companion will enventually be spun up, or encounter internal motion, precisely because of the gravitomagnetic field generated by the companion body. 

Gravitomagnetic tidal interactions were first investigated in the post-Newtonian framework in \cite{Favata:2005da}, and in the relativistic regime in \cite{Damour:Nagar:2009:4, Binnington:Poisson:2009}. However, it has been recently shown that taking into account the internal motion of the fluid driven by the gravitomagnetic perturbation leads to dramatically different values than the gravitomagnetic Love numbers assuming a strict hydrostatic equilibrium \cite{Landry:2015}. It was argued in \cite{Landry:2015} that this assumption of strict equilibrium, i.e. no internal motion, is too restrictive for realistic situations.

The value of the tidal coefficients depends on the equation of state of the fluid. However, it was shown in \cite{Yagi:Yunes:2013:1} that gravitoelectric Love number of neutron stars enjoy universal properties, i.e. they are strongly correlated with the moment of inertia and the rotation induced quadrupole. These universal relations, dubbed I-Love-Q relations involve dimensionless Love number, moment of inertia and quadrupole. This result drew a lot of attention, in particular, the universality of the rotation induced multipoles have been extended to higher multipoles \cite{Yagi:2014bxa, Pappas:2013naa, Chatziioannou:2014tha} leading to an approximate no hair conjecture for neutron stars. The validity of the universal relations has been extended to the case of rapid rotation in \cite{Doneva:etal:2013, Chakrabarti:2013tca}, for frequency corrected Love numbers \cite{Maselli:Cardoso:Ferrari:Gualtieri:Pani:2013} and magnetized neutron stars \cite{Haskell:2013vha}. These uni1versal relations have further been extended some class of alternative models of gravity \cite{Yagi:Yunes:2013:1, Doneva:2014faa, Pani:2014jra, Kleihaus:2014lba, Sham:2013cya}, with the conclusion that they remain undistinguishable from predicted by general relativity, in the regime already constrained by observation, with the exception of the dynamical Chern Simons theory \cite{Alexander:2009tp}, though its viability can be questioned in the non-perturbative regime \cite{Delsate:2014hba}.
These universal relations have been recently used for incorporating tidal effects in waveform simulations in the post-Newtonian framework in \cite{Agathos:2015uaa}.
Importantly, universal relations between various types of Love numbers (shape, electric and magnetic) have been investigated in \cite{Yagi:2013sva}. In particular, it was found that the quadrupole gravitoelectric and gravitomagnetic Love numbers (with strict hydrostatic equilibrium) are correlated with deviation of the order of $10\%$ for realistic equations of state. 

In this paper, we reconsider the universal properties of the quadrupolar gravitomagnetic Love number, under the assumption of internal motion in the star, along the line of \cite{Landry:2015}. In Sec. \ref{sec:model}, we review the model and equations for describing slowly rotating neutron stars and gravitomagnetic perturbation of neutron stars. In Sec. \ref{sec:universal}, we present our results for the gravitomagnetic Love number universality. In particular, we show that a) the gravitomagnetic Love numbers of irrotational bodies are correlated with the rescaled moment of inertia and b) that the universality is enhanced for irrotational bodies, in comparison to strict hydrostatic equilibrium.
We summarize our results in Sec. \ref{sec:ccl}.

\section{Slowly rotating stars and gravitomagnetic Love number}
\label{sec:model}
We use the following line element
\begin{align}
&ds^2 = -b(r)dt^2 + \frac{dr^2}{f(r)} + r^2(d\theta^2 + \sin^2\theta (d\varphi\\
&\quad+\epsilon_r (\omega(r)-\omega_\infty))+ \epsilon_m(h_t(r)  dt + h_r(r) dr)Y_i dx^i,\nonumber
\end{align}
where $\epsilon_r$ and $\epsilon_m$ are respectively the small expansion parameters for slow rotation and tidal deformation, $\omega_\infty$ is the pulsar frequency measured by a distant observer, $Y_i$ is the odd parity vector harmonic, given by $Y_i  = (-\frac{1}{\sin\theta}\partial_\varphi,\sin\theta \partial_\theta) Y_{lm}(\theta,\varphi)$, where $Y_{lm}$ is the standard scalar spherical harmonic, and $x^i = (\theta,\varphi)$.

We model the neutron star fluid by a perfect fluid 
\be
T_{ab} = (\rho + P)u_a u_b + P g_{ab},
\ee
where $g$ is the metric, $\rho$ is the energy density, $P$ is the pressure and $u$ is the four velocity, given by
\be
u^a=\frac{1}{\sqrt{b}}(1,0,0,\epsilon_r (\omega-\omega_\infty))^a
\ee
up to order $\epsilon_r^1$ and $\epsilon_m^0$.

We expand the Einstein equations to first order in $\epsilon_r$ with $\epsilon_m=0$ for slowly rotating neutron stars, and to first order in $\epsilon_m$ with $\epsilon_r=0$ for gravitomagnetic tidal deformations.

The slow rotation equations are given by \cite{Hartle:1967he}:
\bea
&&M' = 4\pi G r^2 \rho,\ b'=\frac{2 b \left(4 \pi G r^3 P+M\right)}{r (r-2 M)},\ P'=-(P+\rho)\frac{ b'}{2 b},\nonumber\\
&&\omega''=\frac{4 \omega' \left(\pi G r^2 (P+\rho )-f\right)}{r f}+\frac{16 \pi G \omega (P+\rho )}{f}.
\eea

We follow \cite{Landry:2015} for computing the gravitomagnetic perturbations equations leading to
\be
r^2h_t'' + \mathcal P h_t'  + V(r) h_t = 0,
\ee
where
\bea
&&\mathcal P = -\frac{4 \pi  r^3 (P(r)+\rho (r))}{f(r)},\\
&&V=\frac{-2 f(r)+8 \pi  \eta  r^2 (P(r)+\rho (r))-(l+2)(l-1)}{f(r)},\nonumber
\eea
where $\eta=-1$ in \cite{Andrade:1999mj, Damour:Nagar:2009}. Recently, Landry and Poisson have shown that $\eta=1$ corresponds to the perturbation of a non rotating fluid with internal motion \cite{Landry:2015}, named irrotational fluid. Here we consider both cases.

In the following, we will work in geometric units, $G=c=1$. The vacuum solution to the slow rotation equations is given by
\begin{equation}
b = f = 1-\frac{2M_s}{r},\ \omega = \omega_\infty \left( 1-\frac{2I}{r^3} \right),
 \end{equation}
where  $I$ is the moment of inertia and $M_s$ the mass of the star.

The vacuum solution of gravitomagnetic perturbation equation is expressed in terms of hypergeometric functions:
\begin{align}
&h_t= \frac{2}{3(l-1)}r^{l+1}\left[ A_4 - 2\frac{(l+1)}{l}K_l^{mag}\frac{2M}{r}^{2l+1}B_4\right]
\end{align}
where 
\bea
A_4 &=& \, _2F_1\left(-l+1,-l-2;-2 l;\frac{2 M}{r}\right),\nonumber\\
B_4 &=& \, _2F_1\left(l-1,l+2;2 l+2;\frac{2 M}{r}\right).
\eea

We define the following dimensionless quantities 
\be
C = \frac{M_s}{R_s},\ \bar I = \frac{I}{M_s^3}, \tilde k_l = K_l (2C)^{2l},
\ee
where $C$ is the compactness, and $R_s$ is the radius of the star. Note that \cite{Landry:2015} defines the compactness as $2M_s/R_s$.

Our goal is to devise a correlation between $\bar I$ and the gravitomagnetic Love number, in particular, we will focus on the case $l=2$. We use tabulated equations of state for describing the interior of the Neutron Star.

\begin{figure}
\includegraphics[scale=.6]{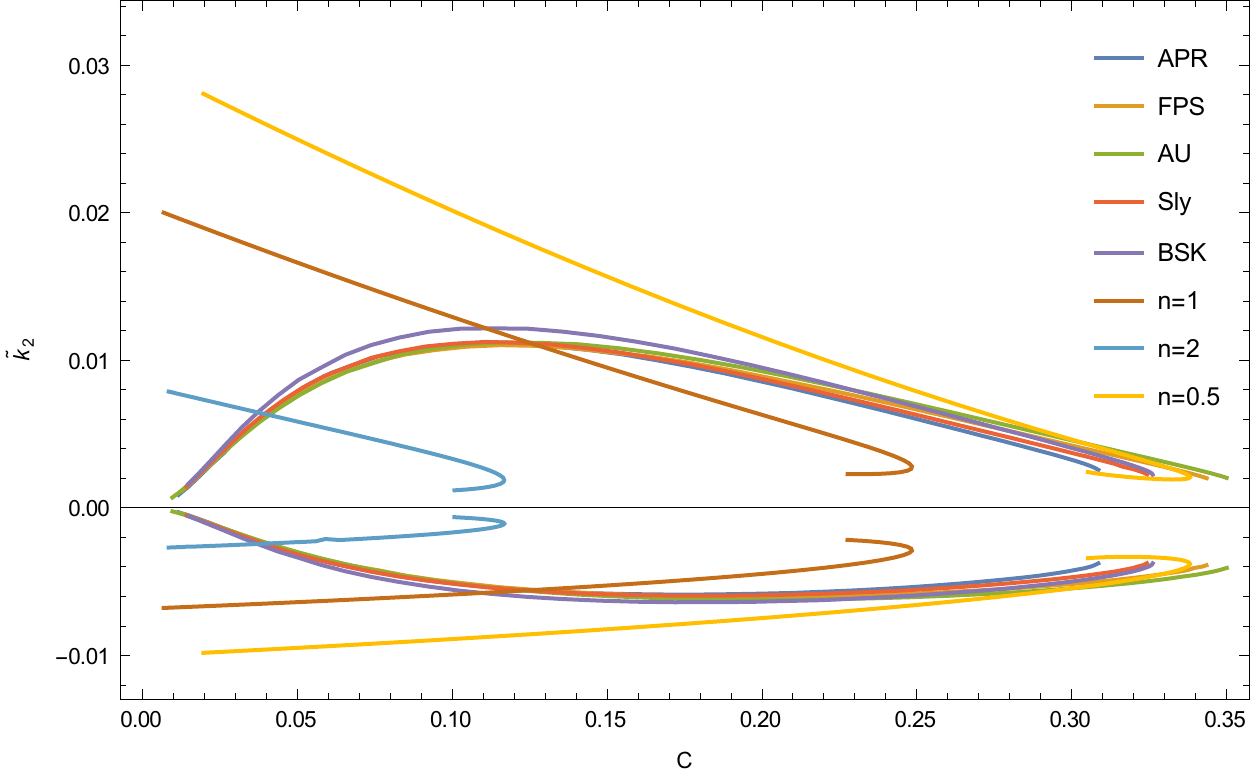}
 \caption{The gravitomagnetic Love numbers for irrotational stars (negative) and strict equilibrium (positive), for realistic tabulated equations of state, and three polytropes, as a function of the compactness.}
 \label{fig:Loves}
\end{figure}

Before doing so, we present for the first time the gravitomagnetic Love numbers of irrotational stars with realistic equations of state in Fig. \ref{fig:Loves}, together with the Love numbers of strict hydrostatic equilibrium.  We note that the gravitomagnetic numbers of irrotational bodies tend to 0 for realistic equations of state. For completeness, we show the gravitomagnetic Love numbers for the polytropic equation of state used in \cite{Landry:2015} for $n=0.5,1,2$; our result agrees perfectly with those of Fig. 1 in \cite{Landry:2015}.

\section{gravitomagnetic Love number universality}
\label{sec:universal}
\begin{figure}
\includegraphics[scale=.4]{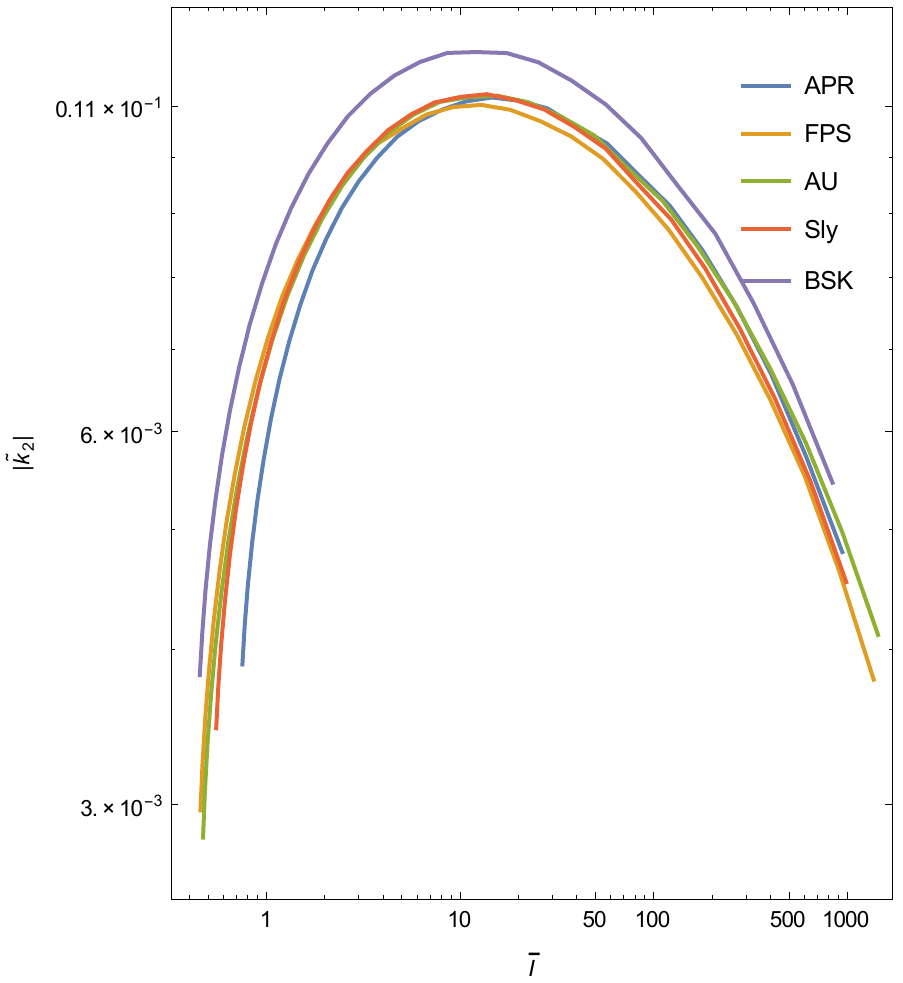}
\includegraphics[scale=.4]{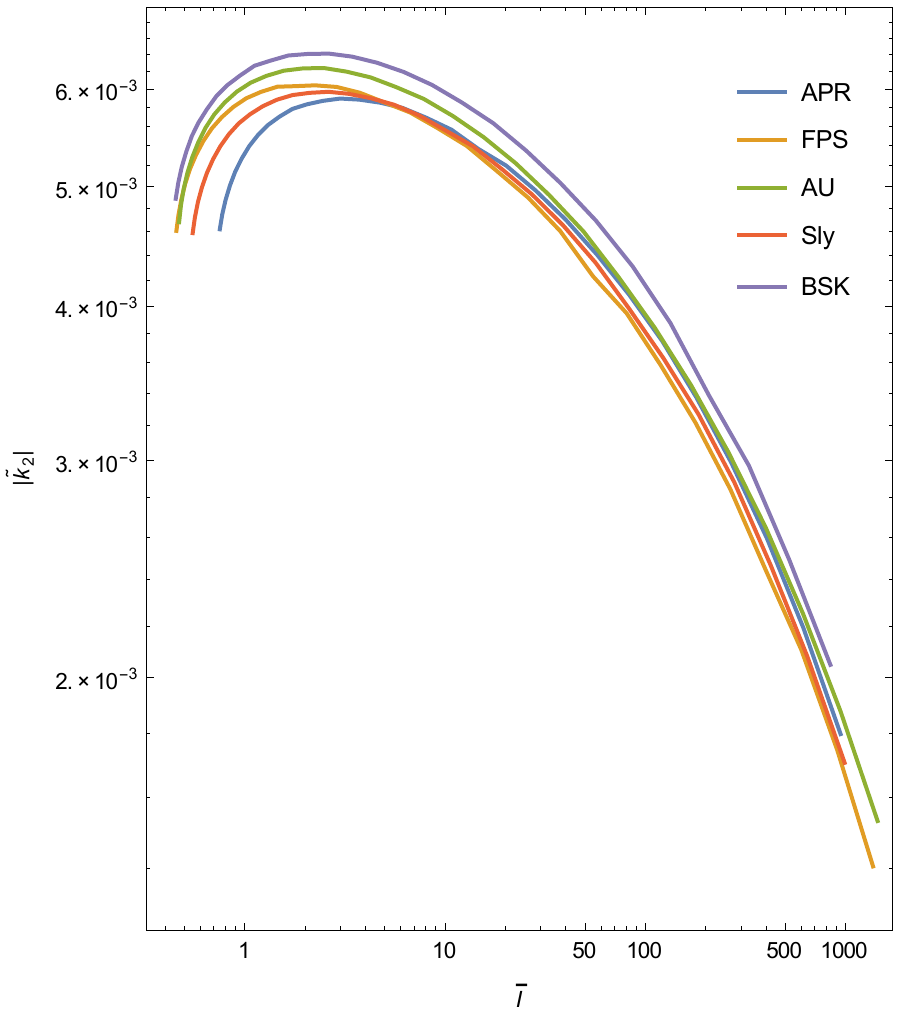}
 \caption{The gravitomagnetic Love number for stars in strict hydrostatic equilibrium (left) and for irrotational stars (right) as a function of the dimensionless moment of inertia (right) in a Log-Log plot, for various equations of state.}
 \label{fig:kI}
\end{figure}

We have integrated the model equations for $5$ representative equations of state, namely APR \cite{Akmal:1998cf}, BSK14 \cite{Goriely:2009zzb}, AU (called AV14+UVII in \cite{Wiringa:1988tp}), SLy \cite{Douchin:2001sv} and FPS \cite{Lorenz:1992zz},  for a range of neutron star mass and radius.
We show the dependence of the gravitomagnetic Love number for strict hydrostatic equilibrium  and for irrotational stars and stars in strict hydrostatic equilibrium in Fig. \ref{fig:kI} as a function of the dimensionless moment of inertia.

Our results indicates that both gravitomagnetic Love number, for the strict equilibrium and for irrotational stars enjoy a quasi universal relations. We find the largest deviation between the best fit and our data for irrotational stars (resp. stars in strict equilibrium) is of the order of $5\%$ (resp. $10\%$), and most of the equations of state are within $\approx2\%$ (resp. $\approx5\%$).
It is in principle possible to reduce this error, along the lines of \cite{Majumder:2015kfa}. We did not find a significant enhancement, but we could reduce the maximum deviation by almost a factor $2$ in the case of strict hydrodynamic equilibrium. Note that the maximal deviation takes into account less compact stars where it is expected that universality deteriorates. In all case, we found universality with less than $\approx 5\%$ for most of the range of compactness that we have investigated. We use the following form of the function to fit the relations
\begin{align}
&\log (C^b \tilde k_2) = d_0 + d_1 \log (C^a \bar I)+ d_2 \log (C^a \bar I)^2+\nonumber\\
&\quad d_3 \log (C^a \bar I)^3+ d_4 \log (C^a \bar I)^4,
\label{eq:fit}
\end{align}
The coefficients are given in Table \ref{tab:fitirr} for irrotational stars and Table \ref{tab:fitstrict} for stars in strict hydrostatic equilibrium.

Note that the deviation is slightly lower for the irrotational case than for the strict hydrostatic equilibrium, which is pleasant feature since this case is supposed to be more realistic.

\begin{figure}
\includegraphics[scale=.8]{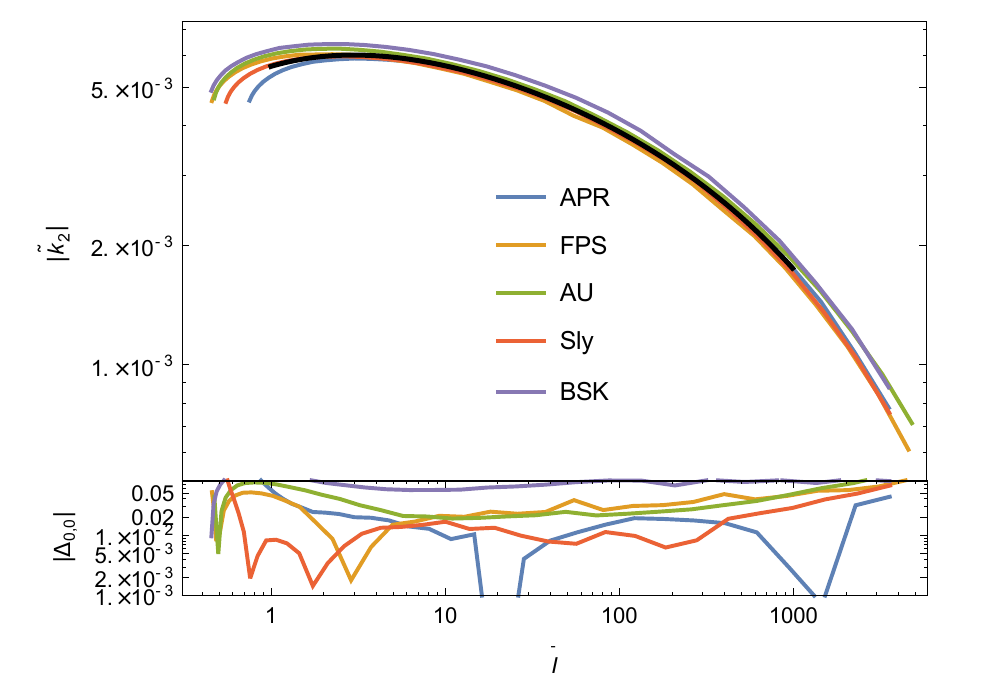}
 \caption{The dispersion fo the gravitomagnetic Love number of irrotational stars The black thick curve is the best fit with parameters given in Table \ref{tab:fitirr}.}
 \label{fig:Dev0Irr}
\end{figure}

\begin{figure}
\includegraphics[scale=.8]{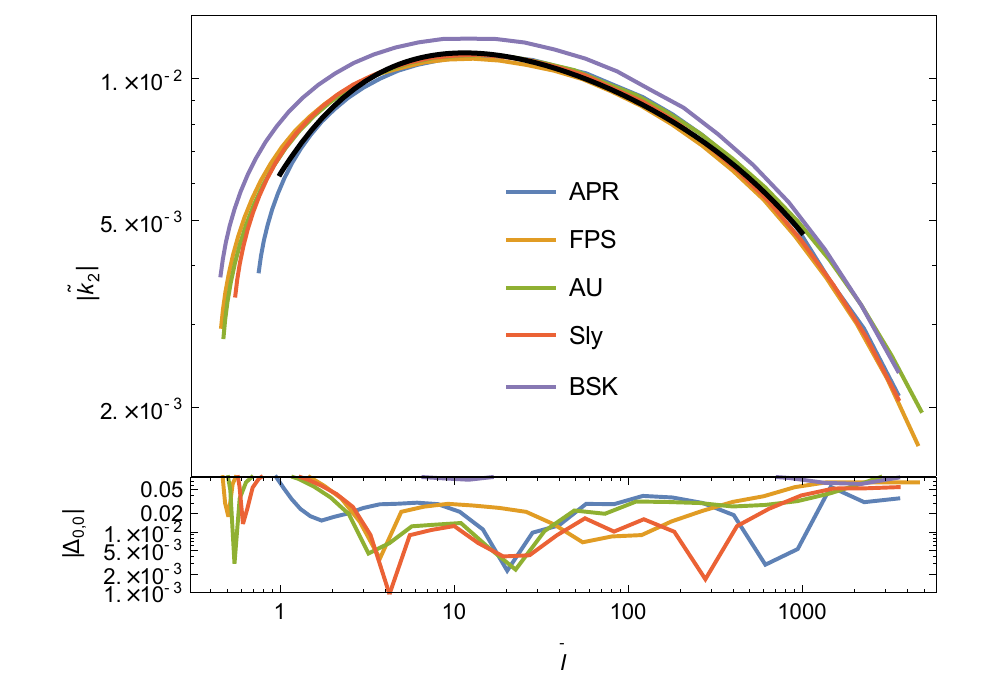}
 \caption{The dispersion fo the gravitomagnetic Love number of stars in strict hydrostatic equilibrium. The black thick curve is the best fit with parameters given in Table \ref{tab:fitstrict}.}
 \label{fig:Dev0Strict}
\end{figure}

We illustrate the deviations for the irrotational case in Fig. \ref{fig:Dev0Irr} and for the strict hydrostatic equilibrium in Fig. \ref{fig:Dev0Strict}. We detail the analysis method we have followed in order to try to improve the universal relations in Appendix \ref{app:bestfit}.

\section{Discussion and conclusion}
\label{sec:ccl}
In this short paper, we have investigate the universality of the gravimagnetic Love number of irrotational stars and stars instrict equilibrium, by looking for correlations between the latter and the dimensionless moment of inertia, defined by $\bar I = I/M_s^3$. We have used $5$ different tabulated equations of state, namely APR, SLy, AU, BSK and FPS, that are commonly used in the literature on universal relations.

We found that both Love numbers normalized follow approximate universal relation, with a dispersion less than $\approx 10\%$ for the strict hydrostatic equilibrium stars and less than $\approx5\%$ for the irrotational stars. Note that previous results were reported on the universal behaviour of the strict hydrostatic equilibrium case in \cite{Yagi:2013sva}. We found that modifying slightly the normalisation of the gravitomagnetic Love number and of the dimensionless moment of inertia according to $C^a \bar I$ and $C^b \tilde k_2$, it is possible to reduce the maximal dispersion by a factor almost $2$ in the case of strict equilibrium. However, the average standard deviation did not change significantly. We gave the best values of $a,b$ that were determined by minimising the total standard deviation, obtained by computing the deviation to the avarage curve of all the data we have generated. We stress that the deviation and mean deviation we computed take into account lighter configurations, where it is expected that the universality holds less.

We expect this universal relation to extend to a the rotation induced quadrupole, as well as to higher gravitomagnetic irrotational multipoles, along with \cite{Yagi:2013sva}. 
Finally, we note that there is a slight improvement in the universality when taking into account the internal motion of the stars, induced by the gravitomagnetic tidal field. We observed than the mean (resp. max) standard deviation in our data is $4\%$ (resp. $14.5\%$) for the irrotational fluid, while it is $7\%$ (resp. $33\%$) in the case of strict static equilibrium. We expect this feature to hold for higher multipoles, strengthening the correlations found in \cite{Yagi:2013sva}.

\section{acknowledgement}
I thank P. Pani, J. Steinhoff and N. G\"urlebeck for useful discussion, and acknowledge partial support from the NewCompStar COST action MP1304.

\appendix
\section{Choosing the normalisation}
\label{app:bestfit}

Following \cite{Majumder:2015kfa}, we investigated a deformation of the univeral relation by looking for correlations between $\bar I C^a$ and  $\tilde k_2 C^b$, for some $a,b$ to be determined.

We adopted the following strategy: first we generated  tables of $(\bar I C^a, \tilde k_2 C^b)$ for all the equations of state we built for various neutron star configurations. Let us call these data $(I_a^K,K_b^K)_i$ and their interpolating function $\mathcal I^K_{ab}(k)$, where $K$ labels the equations of state. Next we have computed the average of the interpolation of these data, and obtained a function 
$ <{\mathcal I}_{ab}(k)>$:
\be
< {\mathcal I}_{ab}(k)> = \frac{1}{N}\sum_{K=1}^N \mathcal I^{K}_{ab}(k),
\ee
where $N$ is the total number of equations of state.

Then the dispersion of the data of the equation of state number $K$ with respect to the avarage values is given by
\be
\Delta_{a,b}^K = \left| \frac{I^K_a -<{\mathcal I}_{ab}(K^K_b)<}{<{\mathcal I}_{ab}(K^K_b)>} \right|.
\ee

It should be stressed that the sampling values of $K_b^K$ depend on $K$. Thus it is not a good idea to use directly these datas as they are. Instead, we resampled the data on an log-equally spaced sample in the common range of the $K^K_a$ datas. This resampling is now independant of the equation of state. Let us call the dispersions computed on the resampled data $D^K_{ab}$. We then compute the variance of the joint data, defining
\be
D_{ab} = \mbox{Var}(\cup_{K=1}^N D^K_{ab}),
\ee
where $\mbox{Var}$ is the variance.

\begin{figure}
\includegraphics[scale=.8]{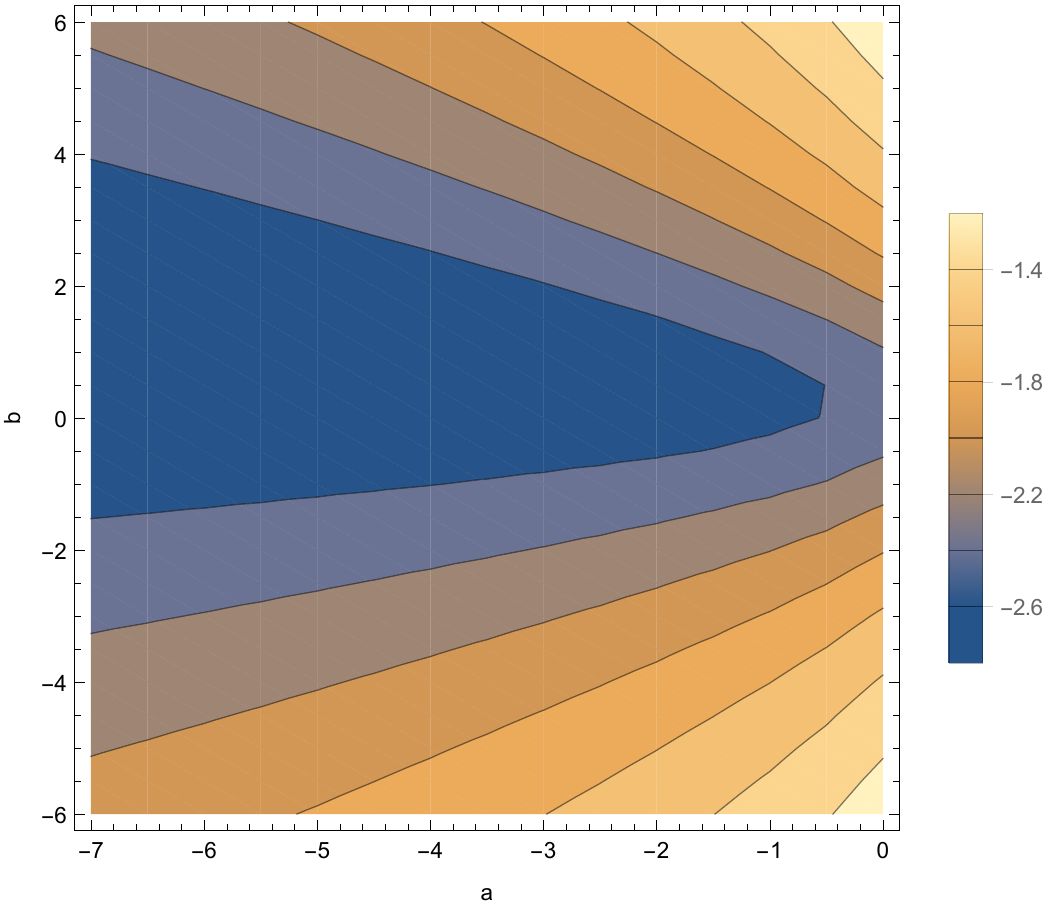}
 \caption{$\log_{10}$ of the cumulative standard deviation of the relative difference between the curves in the $(j_l C^a, \bar I C^b)$ plan and their average, computed with 5 different equations of state for irrotational stars.}
 \label{fig:Var}
\end{figure}

Fig. \ref{fig:Var} shows $\sqrt{D_{ab}}$ as a function of $a$ and $b$. Of course, this plot depends on the specific choice of the equations of state set, but we expect the minimal region to remain roughly the same.
As a sanity check, we checked that the mean value of the dispersion is a small number, typically find of the order of $\approx 10^-3$.

\section{Coefficients of the fit}
 \begin{widetext}
\begin{center}
\begin{table}[h]
\[
\begin{array}{|c|c|ccccc|cc|}
  \hline
  a	&b	&d_0	&d_1	  &d_2	    &d_3	&d_4	   & <|\Delta|> & \mbox{Max}|\Delta|\\
 \hline
 0	&0	&-5.07+0 & 5.96-1& -1.99-1& 2.58-2& -1.51-3 &  0.070     & 0.330\\
 -5     &0	&-8.60+0& 8.69-1& -6.53-2& 2.12-3& -2.79-5  &  0.050	& 0.193\\
 \hline
\end{array}\]    
\caption{Values of the parameter of \eqref{eq:fit}, for the direct relation between $\bar I$ and $\sigma_2$ and for a different choice of normalisation. This case is the case considered in previous literature and refered to as strict hydrostatic equilibrium in \cite{Landry:2015}. $<|\Delta|>$ is the average absolute deviation over all the axis, and $\mbox{Max}|\Delta|$ is largest deviation we observe over all the axis. The numbers are presented as $X\pm Y=X\ 10^{\pm Y}$. }
\label{tab:fitstrict}
\end{table}
%
\begin{table}[h]
\[
\begin{array}{|c|c|ccccc|cc|}
  \hline
  a	&b	&d_0	&d_1	  &d_2	    &d_3	&d_4	   & <|\Delta|> & \mbox{Max}|\Delta|\\
 \hline
 0	&0	&-5.18+0& 1.29-1& -7.16-2& 8.58-3& -6.47-4 &  0.040     & 0.144\\
 -3     &0      &-5.86+0& 2.80-1& -3.38-2& 1.48-3& -3.24-5& 0.034     & 0.131\\
 \hline
\end{array}\]    
\caption{Values of the parameter of \eqref{eq:fit}, for the direct relation between $\bar I$ and $\tilde k_2$ and for a different choice of normalisation. This is the irrotational fluid case considered in  \cite{Landry:2015}, where internal motion is present. $<|\Delta|>$ is the average absolute deviation over all the axis, and $\mbox{Max}|\Delta|$ is largest deviation we observe over all the axis. The numbers are presented as $X\pm Y=X\ 10^{\pm Y}$. }
\label{tab:fitirr}
\end{table}
\end{center}
\end{widetext}

\bibliographystyle{utphys}
\bibliography{maglove}

\end{document}